\documentclass[aps,prd,reprint,twocolumn,superscriptaddress,showpacs,notitlepage]{revtex4-1}

\usepackage{graphicx}
\usepackage{mathrsfs}
\usepackage{bm}
\usepackage{amsmath}
\usepackage{dcolumn}
\usepackage{epstopdf}
\usepackage{dsfont}
\usepackage{amssymb}
\usepackage{tabularx}
\usepackage{array}
\usepackage{float}
\usepackage{color}
\usepackage{epstopdf}
\usepackage{mathrsfs}
\usepackage[colorlinks, linkcolor=blue,anchorcolor=blue,citecolor=blue,urlcolor=blue]{hyperref}
\usepackage[T1]{fontenc}

\begin{document}
\title{Triggering superconductivity, semiconducting states, and ternary valley structure in graphene via functionalization with Si-N layers}

\author{Luo Yan}
\thanks{The first two authors contributed equally to this work.}
\affiliation{School of Physics, University of Electronic Science and Technology of China, Chengdu 610054, China}

\author{Jiaojiao Zhu}
\thanks{The first two authors contributed equally to this work.}
\affiliation{Research Laboratory for Quantum Materials, Singapore University of Technology and Design, Singapore 487372, Singapore}

\author{Bao-Tian Wang}
\affiliation{Institute of High Energy Physics, Chinese Academy of Science (CAS), Beijing 100049, China}

\author{Peng-Fei Liu}
\affiliation{Institute of High Energy Physics, Chinese Academy of Science (CAS), Beijing 100049, China}

\author{Guangzhao Wang}
\affiliation{School of Electronic Information Engineering, Yangtze Normal University, Chongqing 408100, China}

\author{Shengyuan A. Yang}
\email[]{shengyuan\_yang@sutd.edu.sg}
\affiliation{Research Laboratory for Quantum Materials, Singapore University of Technology and Design, Singapore 487372, Singapore}

\author{Liujiang Zhou}
\email[]{liujiang86@gmail.com}
\affiliation{School of Physics, University of Electronic Science and Technology of China, Chengdu 610054, China}
\affiliation{Yangtze Delta Region Institute (Huzhou), University of Electronic Science and Technology of China, Huzhou 313001, China}

\begin{abstract}
Opening a band gap and realizing static valley control have been long sought after in graphene-based two-dimensional (2D) materials. Motivated by the recent success in synthesizing 2D materials passivated by Si-N layers, here, we propose two new graphene-based materials, 2D C$_{2}$SiN and CSiN, via first-principles calculations. Monolayer C$_{2}$SiN is metallic and realizes superconductivity at low temperatures.
Monolayer CSiN enjoys excellent stability and mechanical property. It is a semiconductor with a ternary valley structure for electron carriers. Distinct from existing valleytronic platforms, these valleys can be controlled by applied uniaxial strain. The valley polarization of carriers further manifest as a pronounced change in the anisotropic conductivity, which can be detected in simple electric measurement. The strong interaction effects also lead to large exciton binding energy and enhance the optical absorption in the ultraviolet range. Our work opens a new route to achieve superconductivity, ternary valley structure, and semiconductor with enhanced optical absorption in 2D materials.
\end{abstract}
\maketitle

\section{Introduction}
 Graphene has attracted tremendous research interest in the past twenty years, owing to its excellent electric, mechanical, and optical properties \cite{novoselov2004electric,neto2009electronic}. The study of graphene also boosted the development of the field of valleytronics \cite{schaibley2016valleytronics}.
In graphene, the low-energy carriers are located at two energy degenerate valleys in the momentum space, and it was proposed that this binary valley degree of freedom can be used to encode and process information, analogous to the idea in spintronics \cite{rycerz2007valley,xiao2007valley,yao2008valley}.

In pushing graphene towards electronic applications, a big challenge comes from the absence of a band gap in graphene. Pristine graphene is a semimetal, where the $\pi$ bands, derived from the C-$p_z$ orbitals, cross at two inequivalent Dirac points on the Fermi level, forming the two-valley structure \cite{neto2009electronic}.
Many schemes for opening a band gap in graphene have been proposed. One most direct approach is surface functionalization, i.e., to use the adsorbed atoms or functional groups to saturate the low-energy $p_z$ orbitals. For example, the hydrogenated or fluorinated graphene 2D structures were extensively studied and some successfully demonstrated in experiment as good semiconductors \cite{elias2009control,jeon2011fluorographene}.
Nevertheless, these resulting structures often lacks good thermal/mechanical stability and their carrier mobility is often severely degraded. Moreover, the valley structure of graphene is usually destroyed in the functionalization process.

Very recently, centimeter-scale monolayers (MLs) of the MoSi$_{2}$N$_{4}$ family materials has been successfully synthesized via a novel chemical vapor deposition growth method \cite{hong2020chemical}. 
The structure of MoSi$_{2}$N$_{4}$ can be viewed as a MoN$_{2}$ ML passivated by Si-N layers on its two sides. ML MoN$_{2}$ is metallic; after the Si-N functionalization, the resulting MoSi$_{2}$N$_{4}$ becomes a semiconductor with a band gap of $\sim 1.94$ eV. The similar phase transitions also appear for metallic WN$_{2}$ and CrN$_{2}$ MLs when functionalized by Si-N layers \cite{hong2020chemical,liu2017computational}. 
In addition, it was shown that the Si-N layers can significantly improve the mechanical and thermodynamic stability of these 2D materials, and in some cases lead to interesting physics, such as valley-spin coupling, nontrivial band topology, 2D superconductivity, and piezoelectricity \cite{li2021correlation,li2020valley,wang2021efficient}.

\begin{figure*}[t!]
	\centering
	\includegraphics[width=0.8\linewidth]{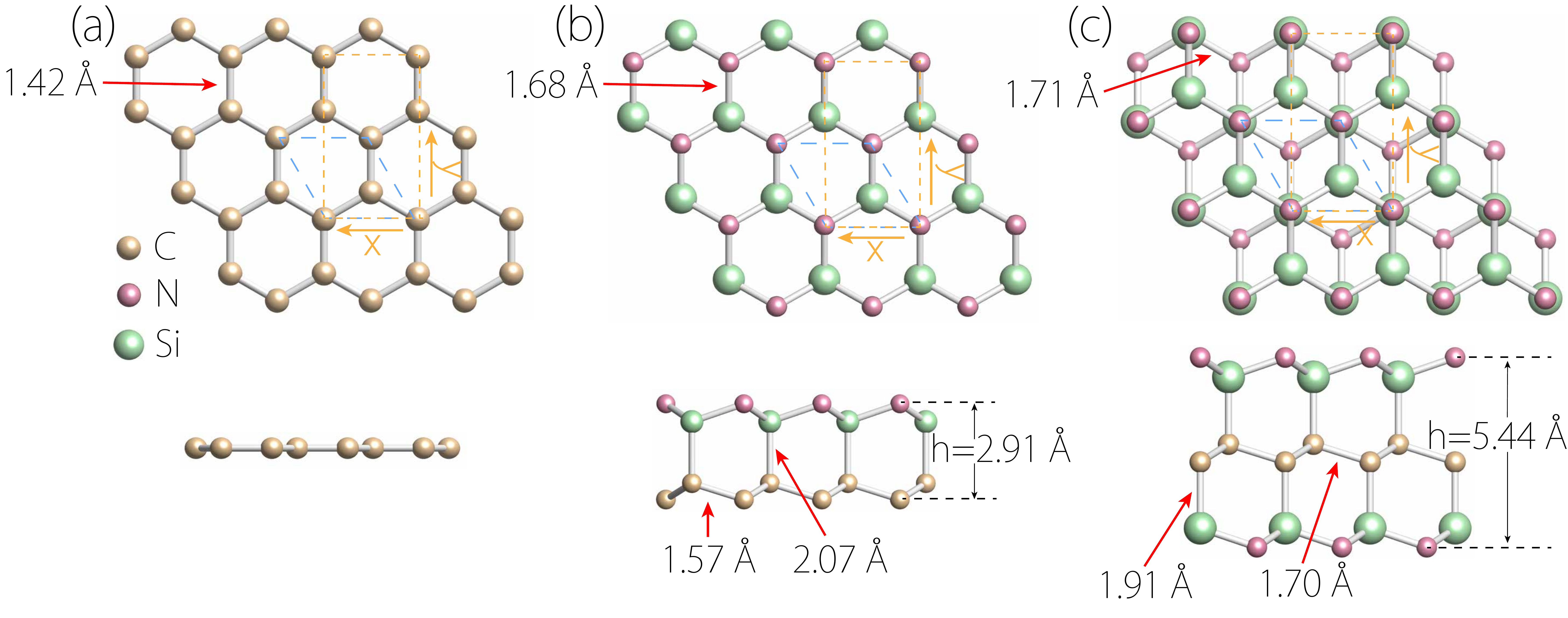}
	\caption{Top (upper panel) and side (lower panel) views for (a) pristine graphene, (b) ML C$_{2}$SiN, and (c) ML CSiN. The blue dashed lines mark the primitive cell, and the orange dashed lines indicate the rectangular conventional cell. }
\end{figure*}

Motivated by the above-mentioned experimental and theoretical progress, in this work, using first-principles calculations, we explore the 2D structures formed by graphene passivated with Si-N layers. Specifically, we consider ML C$_{2}$SiN and CSiN. The former has Si-N layer attached on only one side of graphene, whereas the latter has Si-N layers on both sides. We show that C$_{2}$SiN is a 2D metal and exhibits superconductivity with an estimated $T_c\sim 1.01$ K.
On the other hand,  the 2D CSiN has excellent stability and mechanical property. It is a good semiconductor with band gap $>3$ eV. We find that ML CSiN can maintain a high electron carrier mobility $\sim$ 2000 cm$^{2}$V$^{-1}$S$^{-1}$. Remarkably, 2D CSiN possesses a novel ternary valley structure at the conduction band edge. The three valleys are connected by the $C_{3z}$ symmetry, so the valley splitting and valley polarization in 2D CSiN can be readily controlled by applying an uniaxial strain, which singles out a particular valley label. This is in contrast to the graphene or transition metal dichalcogenides materials, where the valleys are connected by time reversal symmetry and hence the strain control of valley splitting is forbidden. We show that for ML CSiN, a very large valley splitting $>0.6$ eV can be generated at a moderate strain $\sim 5\%$. Importantly, the valley polarization can result in a highly anisotropic electron transport character. In addition, we predict strong excitonic effects in ML CSiN with large exciton binding energy $\sim 1$ eV and strong absorption peak $\sim 5$ eV. Our work provides a new strategy towards graphene-based 2D materials and reveals a new 2D semiconductor platform with novel valleytronic functionalities and excellent mechanical, electronic, and optical performances.

\section{Computational Methods}
Our first-principles calculations are based on the density functional theory (DFT), performed using the Perdew-Burke-Ernzerhof (PBE) functional for the exchange-correlation potential \cite{PhysRevB.50.17953,blochl1994improved},
as implemented in the Vienna \textit{ab initio} simulation package (VASP) \cite{kresse1996efficient,kresse1999ultrasoft}.
The projector augmented wave (PAW) method was adopted to simulate the ionic potentials \cite{blochl1994projector}.
The optB88-vdW approach was used to model the van der Waals interactions \cite{klimes2011van}. 
The kinetic energy cutoff of 450 eV and the \textit{k}-point mesh of 25 $\times$ 25 $\times$ 1 were employed in the calculations. To avoid the artificial interactions between periodic images, a vacuum space of 15 \AA{} was inserted along the \emph{z} direction. The phonon properties were studied within the density functional perturbation theory with PHONOPY code \cite{togo2015first}.
The \textit{ab initio} molecular dynamics (AIMD) simulations with Nos\'{e}-Hoover thermostat \cite{nose1984unified}
were used to evaluate the thermal stability and a 4 $\times$ 4 $\times$ 1 supercell was taken for the simulation. Some data post-processing after VASP calculations was done by using the VASPKIT code \cite{VASPKIT}.
The QUANTUM ESPRESSO (QE) package \cite{giannozzi2009quantum,giannozzi2017advanced}
was used to study the superconductivity within the Bardeen-Cooper-Schrieffer theory \cite{bardeen1957theory}.
To study the excitonic effect, based on the ground-state Kohn-Sham energies and wave functions obtained from QE self-consistent calculations,
YAMBO software \cite{marini2009yambo}
was adopted to model the screened Coulomb interactions (G$_{0}$W$_{0}$ approximation) in combination with the random phase approximation (RPA) or Bethe-Salpeter equation (BSE). In order to converge the quasiparticle energy gap, the total number of bands were set to be 16 times the valence bands and the cutoff energy was set to be 8 Ry for the response function in the  G$_{0}$W$_{0}$ step. The five highest valence bands and five lowest conduction bands were taken to describe the excitons in BSE calculation.

\section{Results and discussion}
\subsection{Crystal structures of ML C$_{2}$SiN and CSiN}
The crystal structures of ML C$_{2}$SiN and CSiN are illustrated in Figure 1b, c. They are constructed by attaching Si-N layers to the ML graphene, similar to the formation of MoSi$_2$N$_4$.
The two crystals share the same space group of \textit{P3m1} (No.~156) with the \textit{C$_{3v}$} point group, which is distinct from the \textit{D}$_{6h}$ point group symmetry for ML graphene (Figure 1a). The optimized lattice parameters for ML C$_{2}$SiN and CSiN are 2.70 and 2.81 {\AA}, respectively. In ML C$_{2}$SiN, the  C$-$C and C$-$Si and Si$-$N bond lengths are found to be 1.57, 2.07 and 1.68 {\AA}, respectively. In ML CSiN, the C$-$C  and Si$-$N bond lengths are elongated to 1.70  and 1.71 {\AA}, whereas the Si$-$C bond length decreases to 1.91 {\AA}. Notably, in both materials, the graphene layer becomes puckered. The puckering height is about 0.2 {\AA} in ML C$_{2}$SiN  and 0.5 {\AA} in ML CSiN. The detailed structural data for the two materials are provided in Supporting Information Table S1 and S2.

To investigate the bonding character, the electron localization function (ELF) is evaluated (Supporting Information Figure S1). The result shows that all bonds in the two structures are of strong covalent type. Particularly, by forming the Si$-$C bonds, the Si atoms saturate the C-$p_z$ orbitals underneath them. The puckered structure of the graphene layer also indicates an evolution from sp$^{2}$ to sp$^{3}$ orbital hybridization for the C atoms.

\subsection{Stability and mechanical property}

\begin{figure}[t!]
	\begin{center}
		\includegraphics[width=0.96\linewidth]{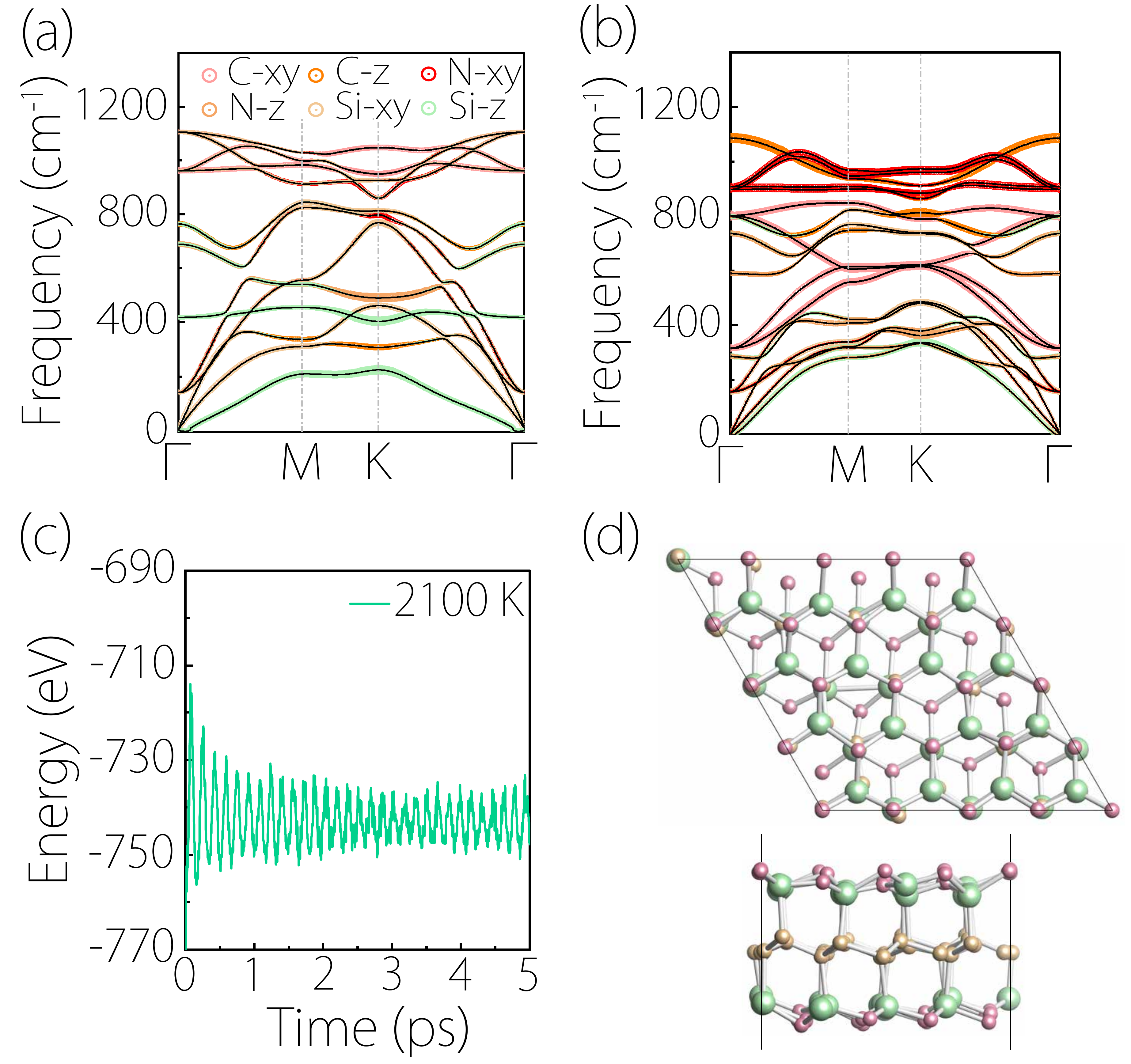}
	\end{center}
	\caption{Phonon spectra of (a) ML C$_{2}$SiN and (b) ML CSiN. The colored weight indicates projection onto vibrational modes of Si, C and N atoms. (c) Energy variation of ML CSiN  during the AIMD simulation at 2100 K. (d) shows the last snapshots from top and side views. }
\end{figure}

The dynamic stabilities of the two materials can be inferred from their phonon spectra, as plotted in Figure 2a, b. The absence of imaginary frequencies in the spectra verifies their dynamic stabilities. One notes that the highest phonon frequency can reach $\sim 1100$ cm$^{-1}$, which is comparable to that of borophene (1200 cm$^{-1}$) \cite{gao2017prediction},
manifesting their strong bonding interactions among the component atoms \cite{song2019two,Yan2020}.
In the vicinity of the $\Gamma$ point, the out-of-plane (ZA) transverse acoustic mode shows a quadratic dependence on the wave vector, while in-plane transverse (TA) and longitudinal  acoustic (LA) modes are in linear dispersions, which is a typical feature for 2D materials. Analysis of the vibration modes shows that the ZA mode mostly involves the out-of-plane vibration of the Si atoms, while the LA and TA modes are related mainly to the in-plane vibration of Si atoms.

We subsequently evaluate the thermal stability of the two materials by performing AIMD simulations. We find that ML C$_{2}$SiN can maintain its structural integrity only around 100 K (Supporting Information Figure S2). In contrast, ML CSiN  is much more stable. Its average value of the energy remains nearly constant with small fluctuations during the entire simulation and no obvious bond breakage at temperature up to 2100 K (Figure 2c, d), showing an extremely high melting point.

Next, we investigate the mechanical properties of ML C$_{2}$SiN and CSiN. For this kind of calculation, it is more convenient to take a rectangle unit cell, as shown in Figure 1. The results show that that the linear elastic regimes for ML  C$_{2}$SiN and CSiN can be up to 5\% strain (Supporting Information Figure S3). Beyond this regime, the plastic deformation arises. The critical strain (i.e., the maximal strain the material can sustain) for the two materials can reach a high value $\sim 20\%$ for both biaxial strain and uniaxial strain along the \textit{y} direction. The critical uniaxial strain along the \textit{x} direction is smaller. It is $\sim$13\% and 9\% for ML C$_{2}$SiN and CSiN, respectively. These results are also supported by their strain-energy curves (Supporting Information Figure S3). The elastic properties of each structure can be characterized by four independent elastic constants: \textit{C}$_{11}$, \textit{C}$_{12}$, \textit{C}$_{22}$ and \textit{C}$_{66}$, which have been evaluated in our calculations. We confirm that they satisfy the Born criterion for the rectangular cell \cite{mouhat2014necessary},
namely, \textit{C}$_{11}$ > 0, \textit{C}$_{66}$ > 0, and \textit{C}$_{11}$ $\times$ \textit{C}$_{22}$ > \textit{C}$_{12}$$^{2}$, indicating that the two materials are mechanically stable. Due to the ambiguity in defining the thickness of a 2D structure, we employ the 2D Young's modulus \textit{Y}$^\mathrm{2D}$ to quantify the in-plane stiffness \cite{zhou2017computational,varjovi2021janus}.
This value is obtained by the relation \textit{Y}$_{x}^\mathrm{2D}$ = (\textit{C}$_{11}$$^{2}$ $-$ \textit{C}$_{12}$$^{2}$) /\textit{C}$_{22}$ and \textit{Y}$_{y}^\mathrm{2D}$ = (\textit{C}$_{11}$$^{2}$ $-$ \textit{C}$_{12}$$^{2}$) /\textit{C}$_{11}$. Using this method, the calculated in-plane stiffness for graphene is 333 N/m, which agrees well with the experimental value of 340 $\pm$ 50 N/m  \cite{lee2008measurement}.
The in-plane stiffness values for ML C$_{2}$SiN and CSiN are calculated to be 382 and 427 N/m, respectively. These values larger than graphene (340 N/m) \cite{lee2008measurement}, h-BN (258 N/m) \cite{topsakal2010elastic}, and MoS$_{2}$ (140 N/m) \cite{peng2013outstanding},
reflecting their strong bonding character. In addition, the Poisson’s ratio ($\nu$) can be obtained from $\nu$$_{x}$ = \textit{C}$_{12}$/\textit{C}$_{22}$ and $\nu$$_{y}$ = \textit{C}$_{12}$/\textit{C}$_{11}$. We find that $\nu$$_{x}$ $\approx$ $\nu$$_{y}$ = 0.19 and 0.26 for ML C$_{2}$SiN and  CSiN, respectively. {More detailed results are given in Supporting Information Table S3.}

\begin{figure*}[t!]
	\begin{center}
	\includegraphics[width=0.9\linewidth]{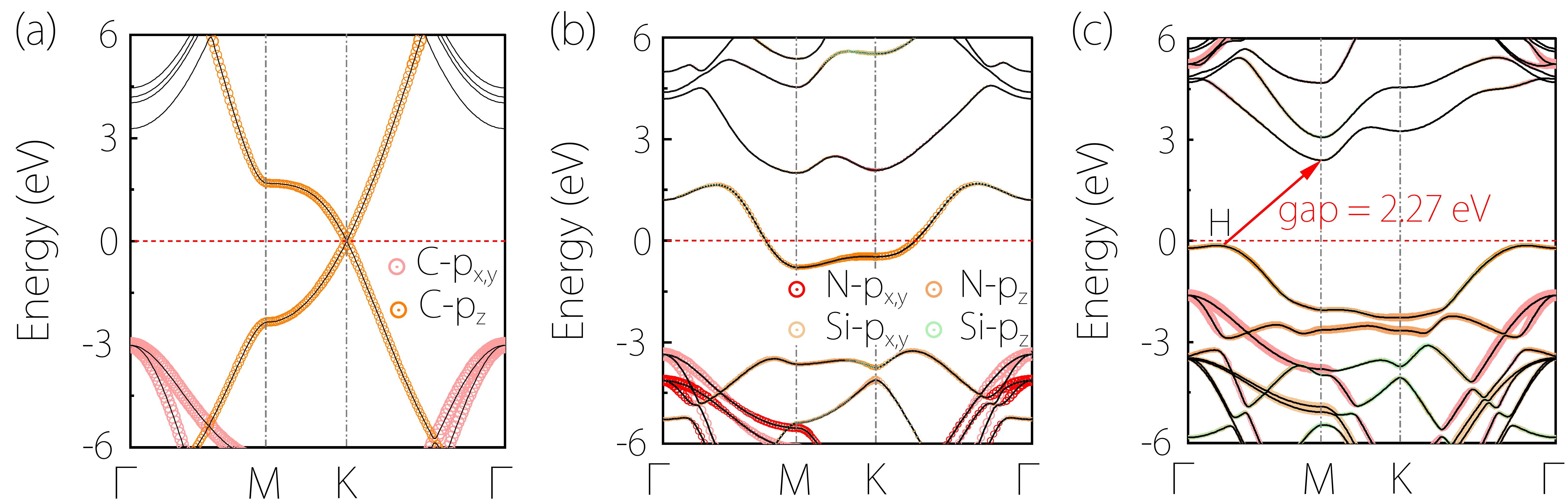}
	\end{center}
	\caption{Band structures of (a) graphene, (b) ML C$_{2}$SiN, and (c) ML CSiN. Projection weights onto different atomic orbitals are indicated. }
	\label{charge}
\end{figure*}

\subsection{Metallicity and superconductivity in ML C$_{2}$SiN}
The orbital projected band structure of ML C$_{2}$SiN  is presented in Figure 3b. For comparison, we also plot the band structure of graphene in Figure 3a. One observes that ML C$_{2}$SiN is metallic, with a single quite flat band crossing the Fermi level. This can be readily understood as following. As discussed, in graphene, the low-energy states are derived from the C-$p_z$ orbitals. In ML C$_{2}$SiN, half of the C atoms in the graphene layer are bonded with the Si atoms in the Si-N layer, with their corresponding $p_z$ orbitals passivated. Nevertheless, there are still another half  C atoms unpassivated. Their $p_z$ orbitals remain at the Fermi level, and the band width is decreased due to the suppressed hopping amplitude. Indeed, the orbital projection in Figure 3b clearly shows that this band is dominated by the $p_z$ orbitals of the unpassivated C atoms, confirming our expectation.

It is known that the pristine graphene is not a superconductor due to its small density of states (DOS) around the Fermi level and its very week electron-phonon coupling strength \cite{cao2018unconventional,neto2009electronic}. Interestingly, we find that ML C$_{2}$SiN can exhibit BCS type superconductivity. We compute the branch magnitude of the electron-phonon coupling (EPC), $\lambda$$_{qv}$, as shown in Figure 4a, which determines the contribution to EPC constant $\lambda$ from individual phonon branch. One can see that the main contribution originates from the low-frequency region (below 450 cm$^{-1}$). In this region, the Eliashberg spectral function $\alpha$$^{2}$F($\omega$) shows two significant peaks, at 200 and 430 cm$^{-1}$, leading to a rapid increase of the cumulative $\lambda$($\omega$), about 75\% of the total EPC ($\lambda$ = 0.32). Physically, the main coupling is between the C-$p_{z}$ orbitals and the out-of-plane vibration modes. Based on the simplified McMillian-Allen-Dynes formula \cite{allen1975transition},
we estimate that the \textit{T$_{c}$} of ML C$_{2}$SiN  is $\sim1.01$ K, which is not high but comparable to the magic-angle-twisted bilayer graphene (1.70 K) \cite{cao2018unconventional}. 
In addition, biaxial strain can further enhance the \textit{T$_{c}$} (to $\sim 7$ K at the 10\% strain), as shown in Figure 4b and S4. Our result demonstrates that superconductivity can be introduced into graphene through surface functionalizations, which opens a new route for achieving superconductivity in 2D graphene-based materials.

\begin{figure}[t!]
	\centering
	\includegraphics[width=1\linewidth]{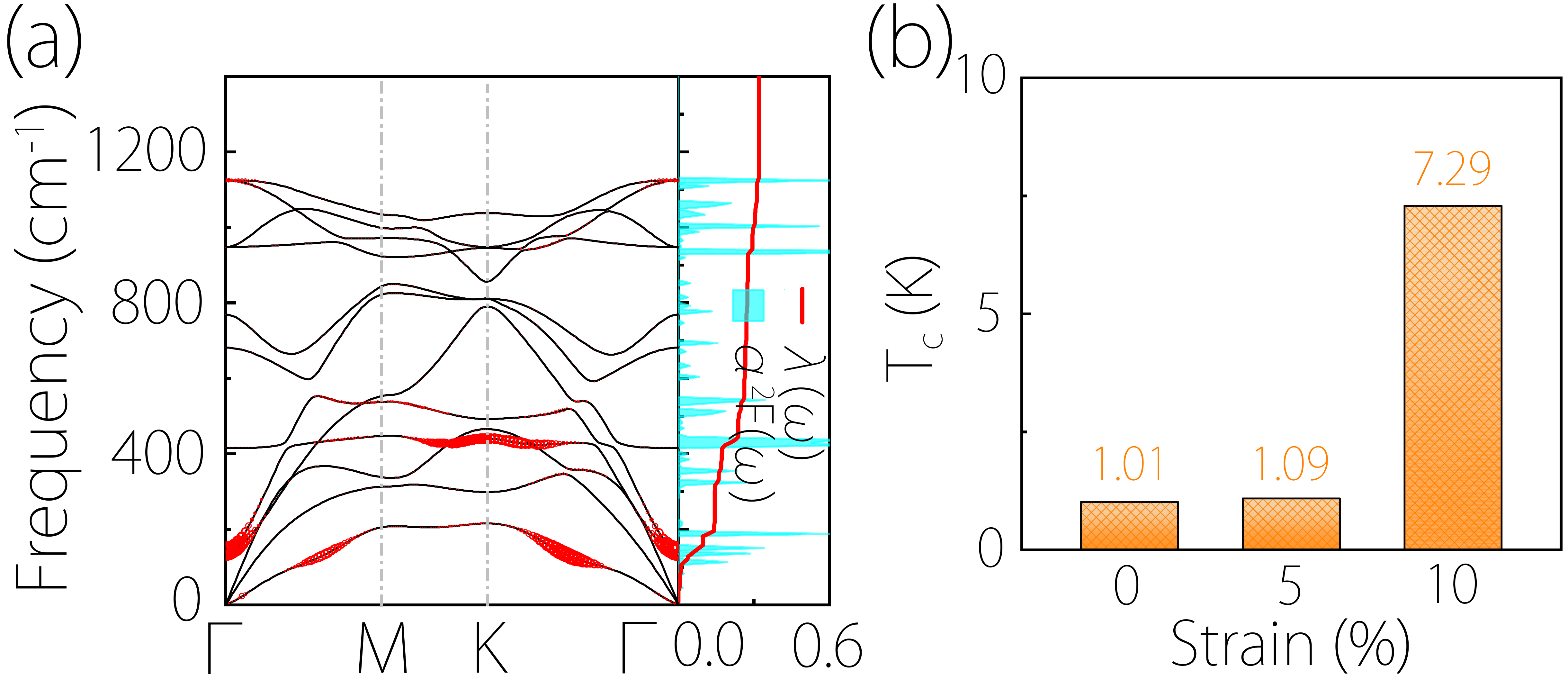}
	\caption{ (a) The weighted EPC $\lambda$$_{qv}$ in the phonon spectrum. The right panel shows Eliashberg function $\alpha$$^{2}$F($\omega$) with cumulative EPC constant $\lambda$($\omega$). (b) Variation of \textit{T$_{c}$} under external strains of 0\%, 5\% and 10\%.}
\end{figure}

\subsection{Ternary valley structure and valley control in ML CSiN}
Now we turn to ML CSiN, which is structurally more robust. Since all the C atoms are bonded with Si atoms in Si-N layers on the two sides, one can expect that all $p_z$ orbitals in the original graphene layer are passivated and the system should become a semiconductor. This picture is confirmed by our result in Figure 3c.
The resulting semiconducting ML CSiN has valence band  maximum (VBM) and conduction band minimum (CBM) locating at $H$ and $M$ points, respectively. Here, $H$ is a point on the $\Gamma$-$M$ path, as indicated in Fig.~3(c). The indirect band gap is about 2.53 eV at the PBE level, and is enlarged to 3.73 eV at the HSE level (Supporting Information Figure S5). From the orbital projection, the valence band is highly dominated by the C-\textit{p}$_{z}$ and N-\textit{p}$_{z}$ orbitals, whereas the conduction band is mostly contributed by Si-\textit{p}$_{x,y}$ orbitals, which are in line with the real-space partial charge density analysis (Supporting Information Figure S6).

\begin{figure}[t!]
	\centering
	\includegraphics[width=0.95\linewidth]{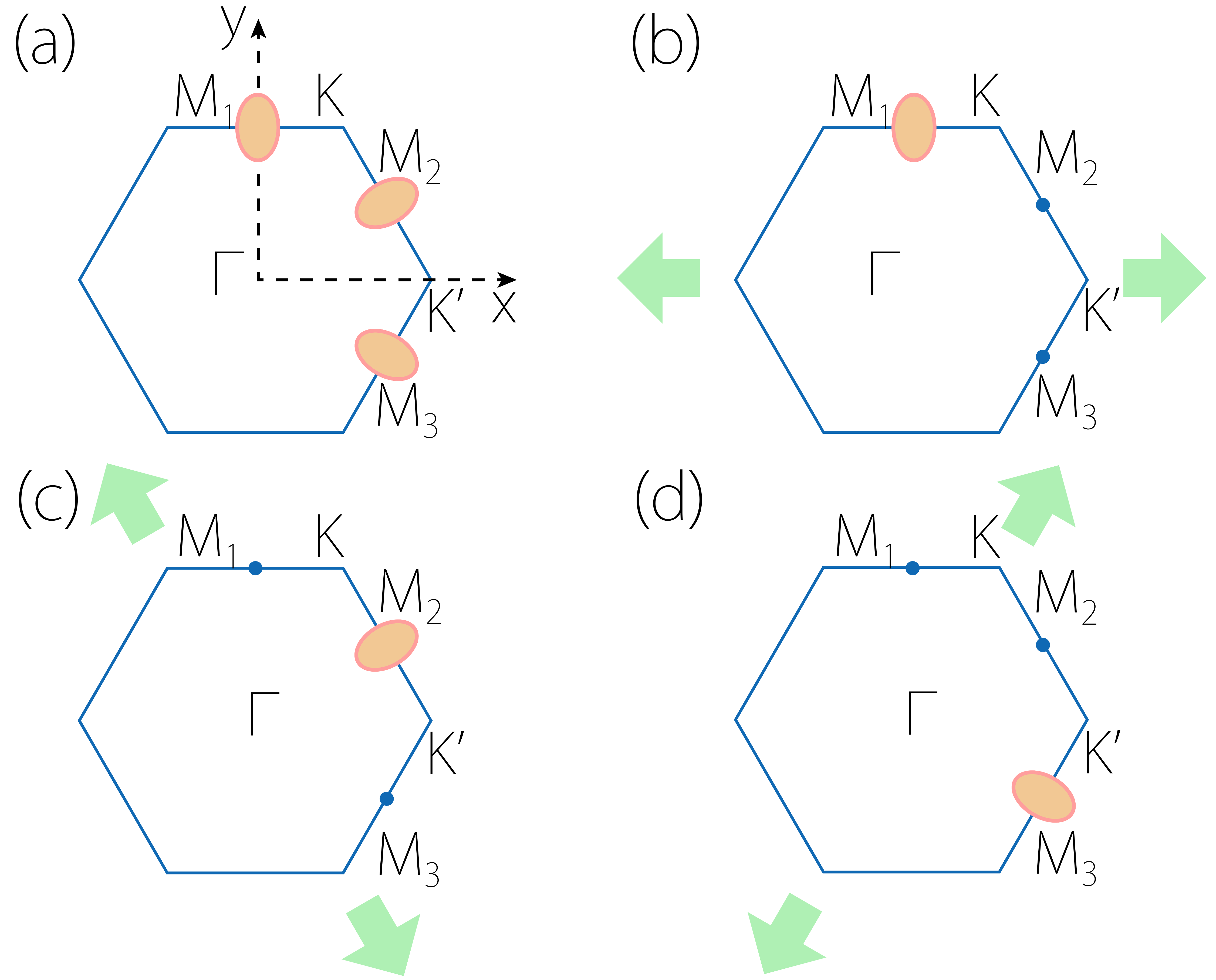}
	\caption{ (a) Schematic figure showing the three $M$ valleys for the electron carriers in ML CSiN. These valleys are degenerate in energy due to the $C_{3z}$ symmetry.  (b) Under a uniaxial tensile strain along $x$, $M_1$ valley shifts down in energy (see Fig.~6(a)). Under small electron doping, all electron carriers are valley polarized in the $M_1$ valley.
(c) and (d) illustrate the cases when the strain is applied along the other two zigzag directions, showing that one can control the valley polarization by strain.}
\end{figure}

\begin{figure}[t!]
	\centering
	\includegraphics[width=0.95\linewidth]{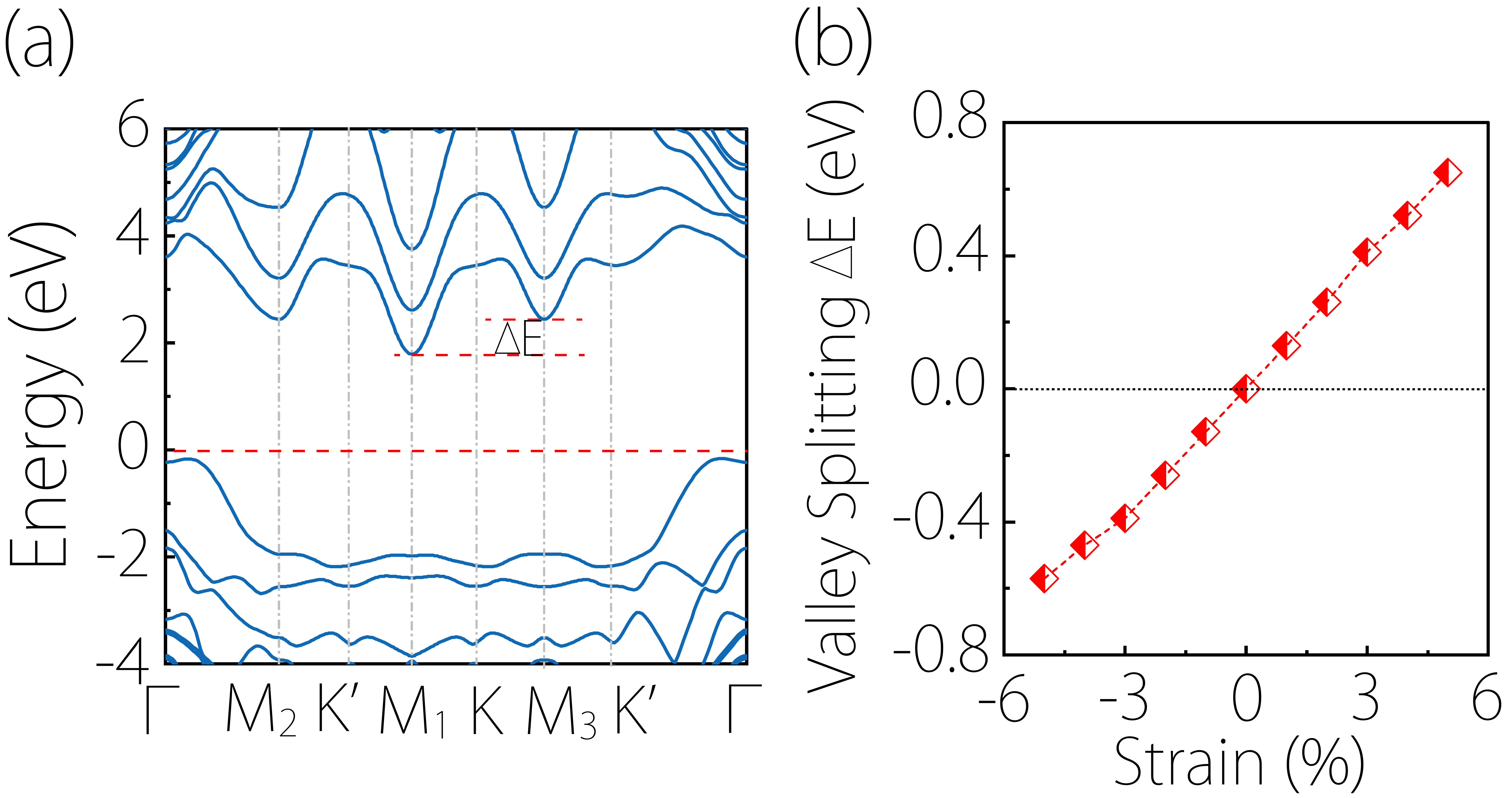}
	\caption{(a) Band structure for ML CSiN at the 5\% tensile strain along the \textit{x} direction. (b) Valley splitting $\Delta E$ as a function of uniaxial strain along \textit{x}.}
\end{figure}

Interestingly, the CBM of ML CSiN exhibits a ternary valley structure, completely different from the binary valley structure in pristine graphene. Consider the electron doped case. As illustrated in Figure 5a, there are three inequivalent valley for the conduction band, centered at the three $M$ points of the Brillouin zone (BZ). The three valleys are connected by the $C_{3z}$ symmetry, which enforces their energy degeneracy. Each electron carrier can be assigned a valley label $M_i$ with $i=1,2,3$.

There is a crucial distinction between ML CSiN and the existing valleytronic platforms, such as graphene or 2D transition metal dichalcogenides \cite{li2021correlation,li2020valley,wang2021efficient,si2013first} . For the latter, the valleys are connected by the time reversal symmetry $\mathcal{T}$. It follows that to generate valley splitting or valley polarization, one must break the $\mathcal{T}$ symmetry, e.g., by using applied magnetic field \cite{cai2013magnetic} or circularly polarized light \cite{yao2008valley} or by non-equilibrium transport \cite{xiao2007valley} . In contrast, for ML CSiN, the three valleys are not connected by $\mathcal{T}$, instead, they are only related by a crystalline symmetry. Therefore, valley splitting and valley polarization in this system can be readily generated in a \emph{static} way, e.g., by using lattice strains that break the crystalline symmetry. Particularly, for 2D materials, strain can be readily applied, e.g., by a beam-bending apparatus or by using piezoelectric substrates \cite{kim2009large,conley2013bandgap}.

The above point is nicely demonstrated in our calculation. As shown in Figure 6a, by applying a uniaxial strain along the \textit{x} direction, the $C_{3z}$ symmetry is broken and the valley $M_{1}$ exhibits an energy splitting from the other two valleys. Figure 6b shows the valley splitting as a function of the applied strain. The valley splitting can be as large as $\Delta E_V= 0.64$ eV at a moderate strain of $\varepsilon= 5\%$. The susceptibility of the valley splitting to strain can be characterized by the slope of the curve:
\begin{equation}
	\chi_V=\left.\frac{d(\Delta E_V)}{d\varepsilon}\right|_{\varepsilon=0}/100,
\end{equation}
which is about 130 meV for ML CSiN.
At small doping level, all the electron carriers will be falling into this $M_{1}$ valley, creating a large valley polarization of the carriers (see Figure 5b). Similarly, by choosing the direction for applying strain, one can control the valley label of the carriers, as illustrated in Figure 5c-d. As discussed, such a valley control scheme is not possible with the conventional valleytronic platforms.

Furthermore, we show that the valley polarization in ML CSiN features a large transport anisotropy, hence it can be detected by electric means. Let's first consider the unstrained case without valley splitting. The mobility of electron carriers can be estimated by the following formula:\cite{bruzzone2011ab,qiao2014few}
\begin{equation}\label{mu}
	\mu_i^\text{2D}=\frac{e\hbar^3C_i^\text{2D}}{k_BT m_d^* m_i^*(D_i)^2},
\end{equation}
where $i$ labels an in-plane direction, $C_i^\text{2D}=(1/S_0)(\partial^2 E_S/\partial \varepsilon_i^2)$ is the 2D elastic constant, $E_S$ and $S_0$ are the energy and area of the system, $\varepsilon_i$ is the strain in the $i$ direction, T is the room temperature (300 K), $m_i^*$ is the effective mass along $i$, $m_d^*=(m_x^* m_y^*)^{1/2}$ is an average effective mass, and $D_i=\partial \Delta/\partial \varepsilon_i$ is the deformation potential (DP) constant, with $\Delta$ the shift of band edge under strain.
Note that without strain, there are three degenerate valleys, so the calculation has to average over the three valleys. Calculated data were given in Table S4. The obtained electron mobility for unstrained ML CSiN is $\mu_x\approx$ 940 cm$^{2}$V$^{-1}$S$^{-1}$, and $\mu_y\approx$ 2391 cm$^{2}$V$^{-1}$S$^{-1}$. Here, $x$ and $y$ directions are indicated in Fig. 1, which are the zigzag and armchair directions, respectively.
These values are larger or comparable to those in MoS$_{2}$ ($\sim$ 200 cm$^{2}$V$^{-1}$s$^{-1}$) \cite{cai2014polarity}
and phosphorene ($\sim$ 1000 cm$^{2}$V$^{-1}$s$^{-1}$) \cite{liu2014phosphorene}.

Now we consider a $5\%$ uniaxial strain applied in the \textit{x} direction. As shown in Figure 6a, the CBM lies in only one valley, the $M_{1}$ valley, and at low doping, the electrons are fully valley-polarized in this valley. Importantly, the individual valley is highly anisotropic (the little co-group at $M$ point is $\textit{C}_{s}$), which can be readily seen from the effective masses. For the $M_{1}$ valley, we find $m_x^*=0.26 m_{e}$ and $m_y^* =0.81 m_{e}$ By using the formula (\ref{mu}), we obtain that when the carriers are valley-polarized in the $M_{1}$ valley, the mobility becomes $\mu_x$= 1388 cm$^{2}$V$^{-1}$S$^{-1}$ and $\mu_y$= 270 cm$^{2}$V$^{-1}$S$^{-1}$. One observes that the ratio $\mu_x/\mu_y$ changes by about 10 times from $\sim 1/2$ at unstrained case to $\sim 5$ at $5\%$ strain.
This can be readily detected as an anisotropy in the conductivity $\sigma_x/\sigma_y\approx \mu_x/\mu_y$.

Therefore, in ML CSiN, we can use moderate strain to induce highly anisotropic electron transport. For valleytronic applications, the system offers a novel ternary valley structure. The valley splitting and polarization can be readily controlled by strain in a static manner. And the valley polarization of carriers (i.e., the valley information) can be further read out in a fully electric way by measuring the anisotropy in conductivity/resistivity.

\begin{figure}[t!]
	\begin{center}
		\includegraphics[width=1.0\linewidth]{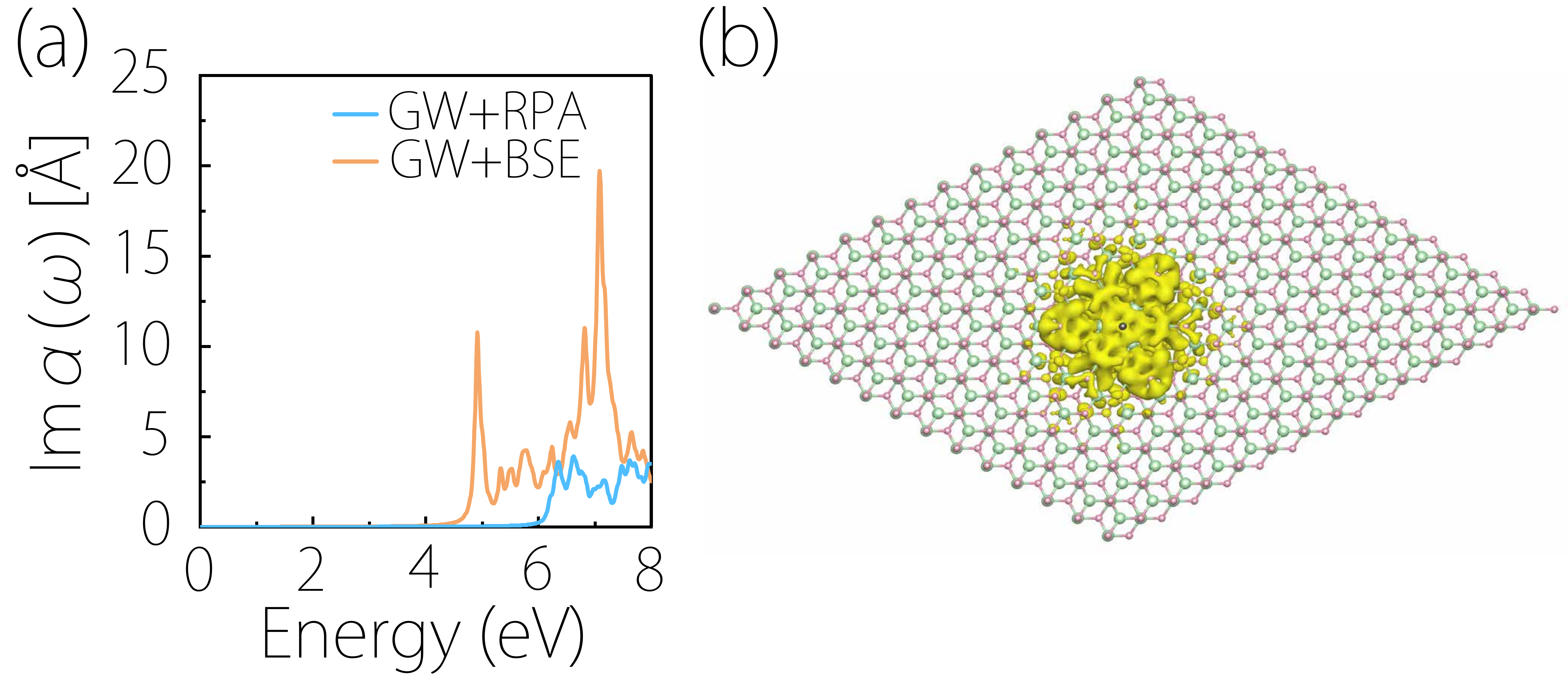}
	\end{center}
	\caption{(a) Imaginary part of the transverse dielectric function of ML CSiN, calculated at the level of G$_{0}$W$_{0}$+RPA and G$_{0}$W$_{0}$+BSE. (b) Spatial distribution of the first bright exciton (marked by the arrow in (a)). Here, we use a $13\times 13 \times 1$ supercell. The hole position is marked by the black spot at the center. The isosurface value is set to be 0.03 e/{\AA}$^{3}$.}
\end{figure}

\subsection{Excitonic effect in ML CSiN}

We have shown that ML CSiN is a good semiconductor. Typically, 2D semiconductors have strong excitonic effect due to the reduced screening of Coulomb interactions \cite{zhou2017computational,yan2022two}.  To capture this effect, we adopt the GW approximation based on the quantum many-body perturbation theory \cite{hedin1965new}
to explore the optical properties.
The frequency-dependent 2D polarizability of the material can be obtained from \cite{rasmussen2016efficient}
\begin{align}
	\alpha_\text{2D}(\omega)=-\lim_{\textbf{q}\longrightarrow0}\dfrac{\chi(\textbf{q}, \omega)}{\textbf{q}^{2}}
\end{align}
where \textbf{q} is the  in-plane wave vector and $\chi$ is the polarizability function.
The imaginary part of $\alpha_\text{2D}(\omega)$ reflects the optical absorption property of the material. For ML CSiN, the results are plotted in Figure 7a. Here, we have taken two approaches: one is GW plus random phase approximation (GW+RPA) [without electrn-hole (e-h) interactions] and the other is GW plus Bethe-Salpeter equation (GW+BSE) (with e-h interactions). Their difference can manifest the exciton contribution.

One observes that when e-h interactions are taken into consideration, a sharp exciton absorption peak emerges at about 4.91 eV (the optical gap). According to the spectral edge from GW+RPA calculation, the G$_{0}$W$_{0}$ gap is about 6.0 eV. Then, the exciton binding energy is estimated to be about 1.09 eV and is mainly associated to direct transitions at $\Gamma$ point. {This binding energy is larger than that in MoS$_{2}$ (0.80 eV) \cite{jin2018prediction},
ML Janus-MoSSe  (0.95 eV) \cite{long2021effect}
and GeSe (0.40 eV)  \cite{gomes2016strongly},
due to its larger bandgap with weaker screening.} The spatial extent of the exciton state can be studied by plotting the magnitude of the exciton wave function given by
\begin{align}
	|\psi (\textbf{r}_{e}, \textbf{r}_{h})| = \sum_{c\nu\textbf{k}}A_{c\nu\textbf{k}}^S\psi_{c\textbf{k}}(\textbf{r}_{e})\psi_{\nu\textbf{k}}(\textbf{r}_{h}),
\end{align}
where $A_{c\nu\textbf{k}}^S$ is the e-h amplitude, and \textbf{r}$_{e}$ and \textbf{r}$_{h}$ denote the electron and hole coordinates, respectively. In Figure 7d, we plot the spatial distribution for the first exciton peak. One clearly sees that the exciton is tightly bonded with a narrow radius $\sim$ 5.83 {\AA}, consistent with its strong binding energy. From Figure 7a, we can conclude that ML CSiN exhibits strong excitonic effect, which greatly enhances its optical absorption in the ultraviolet range.

\section{CONCLUSIONS}
In conclusion, we propose two graphene-based 2D materials, the ML C$_{2}$SiN and CSiN, motivated by the recent experimental progress on 2D structures passivated with Si-N layers. We find that C$_{2}$SiN is a metal and exhibits superconductivity unavailable in pristine graphene.
The ML CSiN has excellent stability and mechanical property. It is an indirect gap semiconductor with band gap $>3$ eV, and its electron mobility can reach $\sim$2000 cm$^{2}$V$^{-1}$S$^{-1}$. Importantly, ML CSiN has a ternary valley structure for electron carriers. In contrast to existing valleytronic platforms, the valleys in ML CSiN are connected by a crystalline symmetry instead of the time reversal symmetry. This enables a static control of valley polarization in ML CSiN, e.g., by uniaxial strain. We show that the valley polarization can be readily detected via purely electric measurement as an anisotropy in the conductivity. Finally, we show strong excitonic effects in ML CSiN with large exciton binding energy $\sim 1$ eV and strong absorption peak in the ultraviolet range. Our work reveals a novel type of 2D valleytronic platform with new valley control and detection mechanisms. Based on their excellent properties, the two new materials could also find useful applications in mechanical, electronic and optical devices.

\section*{ACKNOWLEDGEMENTS}
The authors thank D. L. Deng for valuable discussions. This work is supported by the Startup funds of Outstanding Talents of UESTC (A1098531023601205), National Youth Talents Plan of China (G05QNQR049), the Open-Foundation of Key Laboratory of Laser Device Technology, China North Industries Group Corporation Limited (KLLDT202106), and Singapore MOE AcRF Tier 2 (MOE-T2EP50220-0011).  B.-T.W. acknowledge financial support from the Natural Science Foundation of China (Grants No. 11675195 and No. 12074381) and Guangdong Basic and Applied Basic Research Foundation (Grant No. 2021A1515110587).

\bibliographystyle{apsrev4-1}
\bibliography{apssamp}
\end{document}